# Efficient spin-orbit torque driven magnetization switching of GdFe using phosphorus-implanted platinum layers

Kazuki Shintaku[1], Arun Jacob Mathew[1], Akihisa Iwamoto[1], Mojtaba Mohammadi[2], Hiroyuki Awano[2], Hironori Asada[3], Yasuhiro Fukuma[1, 4*]

[1]Department of Physics and Information Technology, Faculty of Computer Science and Systems Engineering, Kyushu Institute of Technology, Iizuka 820-8502, Japan

[2]Memory Engineering Laboratory, Toyota Technological Institute, Nagoya 468-8511, Japan

[3]Graduate School of Sciences and Technology for Innovation, Yamaguchi University, Ube 755-8611, Japan

[4]Research Center for Neuromorphic AI hardware, Kyushu Institute of Technology, Kitakyushu 808-0196, Japan

**Abstract:**

The capability of the spin-orbit torque (SOT) generated via phenomena such as the spin Hall effect in heavy metals, in switching the magnetization of an adjacent magnetic material, has been studied extensively over the last decade. The efficiency of SOT generation is commonly quantified in terms of the spin Hall angle $\theta_{SH}$. In this work, we demonstrate experimentally that implanting platinum (Pt) with phosphorus (P), resulting in Pt (P) $d$, where $d$ denotes the implantation dose, increases $\theta_{SH}$ by a factor of 7, from 0.06 ($d$ = 0) to 0.43 ($d = 10 \times 10^{16}$ ions/cm$^2$). The enhanced $\theta_{SH}$, along with factors such as perpendicular magnetic anisotropy and resistivity, lead to reduction of the critical current density for switching the perpendicular magnetization of ferrimagnetic rare earth-transition metal alloy Gd$_{26}$Fe$_{74}$, by a factor of nearly 27, from $4.0 \times 10^{11}$ A/m$^2$ ($d$ = 0) to $1.5 \times 10^{10}$ A/m$^2$ ($d = 10 \times 10^{16}$ ions/cm$^2$). Further, the switching current density at zero thermal fluctuations and thermal stability factor were evaluated and found to be $2.0 \times 10^{10}$ A/m$^2$ and 61.4 ($d = 10 \times 10^{16}$ ions/cm$^2$), with the latter being sufficiently above the required threshold for commercial memory applications. Our results suggest that Pt (P) could be a strong candidate in realizing efficient SOT driven magnetization switching leading to the development of improved memory and logic devices in the future.

[*]Author to whom correspondence should be addressed: fukuma@phys.kyutech.ac.jp

The ability to control the magnetization of magnetic materials via electrical means has been pursued with rigorous interest over the last few decades, especially following the discovery of spin transfer torques (STTs) close to the turn of the millennium.[1,2] These persistent efforts eventually led to the commercialization of STT magnetic random access memories (MRAMs).[3] In the STT mechanism, the current flowing through a fixed ferromagnetic (FM) layer gets spin-polarized. This spin-polarized current proceeds to flow into a neighboring free FM layer transferring angular momentum to the free layer magnetization, thus resulting in the action of a torque.[4] However, STT driven magnetization switching presents several notable challenges. Passing high switching currents over multiple cycles increases the probability of device breakdown.[5] Further, since the direction of spin polarization is parallel to the fixed FM layer, the free layer magnetization should necessarily be tilted away from being aligned exactly parallel or antiparallel to the fixed layer magnetization for STT to act upon it.[4] This necessitates thermal fluctuations be acting on the free layer magnetization to promote misalignment, thereby making the switching process slower.

More recently, an alternative current-driven spin torque which arises due to the spin-orbit coupling (SOC) in nonmagnetic (NM) heavy metals, has gained significant attention.[6,7] The nature of its origin has led to this torque being referred to as the spin-orbit torque (SOT).[8] SOT offers solutions to several of the challenges faced in STT driven switching.[9] The high switching currents pass through only the NM layer, resulting in better device endurance. In conventional SOT driven perpendicular magnetization switching, since the spin polarization is orthogonal to the magnetization, the switching dynamics has been found to be faster as compared to STT driven switching.[10] These potential advantages have led to SOT being widely regarded as holding great promise in realizing the next generation of memory and computing technologies.[11–13]

The physical mechanisms responsible for the generation of SOT may predominantly be interfacial SOC effects such as the Rashba-Edelstein effect[14] or bulk SOC effects such as the spin Hall effect (SHE),[15] depending upon the thickness of the NM. In bulk NMs with SHE as the dominant mechanism of SOT generation, passing an in-plane charge current density $J_c$ through the NM results in a pure spin current density $J_s$ flowing out-of-plane.[7] The interconversion efficiency between $J_c$ and $J_s$, is quantified in terms of the spin Hall angle $\theta_{SH}$,[15] and given by $\vec{J}_s = \left(\frac{\hbar}{2e}\right)(\vec{\sigma} \times \vec{J}_c)\theta_{SH}$, where $\sigma$, $\hbar$ and $e$ denote the spin polarization, reduced Plack's constant and electronic charge, respectively. Enhancing $\theta_{SH}$ is of great significance, from both fundamental physics and technological points of view.

The first attempts in utilizing SOT to realize perpendicular magnetization switching were performed using conventional NM/FM layers such as Pt/Co,[6,7] Ta/CoFeB,[16,17] etc. possessing interfacial perpendicular magnetic anisotropy (PMA). In comparison to conventional FMs, ferrimagnets (FiMs) having unequal, antiparallel sublattices, offer bulk PMA, flexibility in magnetic properties as a function of temperature and/or composition, reduced stray fields and faster magnetization dynamics.[18] These advantages have led to extensive investigations of SOT driven switching in NM/FiM systems at room temperature.[19,20] Further, an increase in $\theta_{SH}$ close to magnetization compensation[21,22] as well as faster domain wall motion near angular momentum compensation[23–26] have been reported in various FiMs.

There have also been attempts to improve $\theta_{SH}$ by modifying the NM layer. These attempts include implantation of the NM with lighter elements,[27–29] doping the NM with lighter elements during deposition,[30] etc. Novel materials such as two dimensional electron gases,[31] transition metal dichalcogenides[32] and topological insulators (TIs)[33] have also been investigated in this regard. TI/FiM samples have been found to exhibit switching current densities an order smaller as compared to conventional NM/FiM structures.[34] However, complete switching could not be demonstrated using this enhanced $\theta_{SH}$. Further, in addition to $\theta_{SH}$, factors such as the perpendicular anisotropy field $H_p$ and thermal stability factor $\Delta$ need to be evaluated in order to ascertain the overall viability of a proposed material having enhanced $\theta_{SH}$ for use in perpendicular magnetization switching.

In this letter, we utilize Pt implanted with P ions accelerated at a potential of 30keV, referred to henceforth as Pt (P), as the SOT source. Complete switching of the perpendicular magnetization of ferrimagnetic rare earth-transition metal alloy $Gd_{26}Fe_{74}$ is performed at decreasing current densities with increase in P implantation dose $d$ (expressed in units of ions/cm$^2$). The contribution of PMA and $\theta_{SH}$ on the critical switching current densities $J_{sw}$ are studied in detail, via anomalous Hall measurements[35] and harmonic Hall measurements,[36,37] respectively. For nonzero $d$, $H_p$ is found to decrease while $\theta_{SH}$ increases. Both these factors, in conjunction, lead to reduced $J_{sw}$ at higher $d$. We proceed to confirm that the reduced $H_p$ does not lead to undesirable thermal fluctuations of the perpendicular magnetization by evaluating $\Delta$ for nonzero $d$. In this fashion, we perform a comprehensive evaluation of complete magnetization switching of the GdFe ferrimagnet driven by enhanced SOT generated by Pt (P), while taking into due consideration, multiple factors that could potentially contribute to deciding the feasibility of efficient current-driven switching.

We began by sputtering Pt (10)/MgO (10)/Al$_2$O$_3$ (10 nm) on Si/SiO$_2$ substrates. These multilayers were subsequently implanted with different doses of P under an ion beam having an accelerating voltage of 30keV. After implantation, we renamed the stacks as Pt (P) $d$ (10)/ MgO (10)/Al$_2$O$_3$ (10 nm) where $d$ denotes dose of P ion implantation (in units of ions/cm$^2$). The results reported in this work are based on measurements performed upon samples having three different $d$ values, namely 0, $5 \times 10^{16}$ and $10 \times 10^{16}$. Following this, the oxide layers were milled out by argon ion milling, and photoresist was patterned over the samples by ultraviolet photolithography. Pt (1)/Gd$_{26}$Fe$_{74}$ (20)/SiN (10 nm) layers were sputtered on the patterned samples. The FiM was deposited by room temperature co-sputtering of Gd and Fe targets. Following lift-off, Ti (10)/Al (200 nm) electrodes were made on the samples for performing electrical measurements on the fabricated microdevices.

The magnetic anisotropy of the samples was characterized by performing anomalous Hall measurements based on the schematic measurement setup shown in Fig. 1(a) . A dc read current $I_{\text{read}} = + 1$ mA was applied along $x$, and Hall voltage $V_{\text{Hall}}$ was measured while sweeping the external magnetic field $H$. Figure 1(b) shows normalized $V_{Hall}$ as a function of $H$ swept in the out-of-plane direction along $z$. Irrespective of $d$, square hysteresis loops were obtained that indicated PMA. In other words, when all external stimuli (magnetic field, current, etc.) are removed, the magnetization prefers to align in the out-of-plane orientation. The maximum (minimum) value of $V_{Hall}$ corresponds to when magnetization is oriented entirely along $+z$ ($-z$). $H_c$ is defined as the field $H_z$ at which the polarity of $V_{Hall}$ reverses. The polarity of the obtained signal indicates that GdFe, in all our samples, is Fe-dominated.

Figure 1(c) depicts normalized $V_{Hall}$ when $H$ is swept in the in-plane direction, along $x$. Irrespective of $d$, $V_{Hall}$ is found to tend to zero with increase in $H_x$. We perform a linear fit on the signals in Fig. 1(c), as indicated by the solid lines. The x-intercept obtained gives a measure of $H_p$, i.e. the value of $H_x$ at which $V_{Hall}$ is zero. A comparison of $H_c$ and $H_p$ for different doses is shown in Fig. 1(d). Both $H_c$ and $H_p$ are observed to reduce for finite $d$ as compared to when $d = 0$. This indicates that the layers below the FiM significantly affect its magnetic anisotropy.

The harmonic Hall technique was used to evaluate $\theta_{SH}$. A sinusoidal alternating current ($I_{ac}$) having frequency 79 Hz, was given as input and the first ($V_{1f}$) and second ($V_{2f}$) harmonic Hall voltages and were measured using a lock-in amplifier, under an $H_x$ sweep. $V_{1f}$ exhibits $H_x$ dependence similar to the previously measured dc measurement, and hence can be fitted to a parabola given by,

$$V_{1f}(H_x) = a + bH_x + cH_x^2 \tag{1},$$

where $a$, $b$ and $c$ are fitting parameters. Figures 2(a) and 2(b) depict $V_{1f}$ under $H_x$ sweep for $d = 0$ and $d = 10 \times 10^{16}$ ions/cm², respectively, for different $I_{ac}$. With increase in $I_{ac}$, the magnitude of $c$, i.e. the curvature of the fitted parabola, is also found to increase. The contribution of SOT is contained in $V_{2f}$. For external in-plane fields significantly smaller than $H_p$, $V_{2f}$ can be expressed as a linear function of $H_x$, as given by,

$$V_{2f}(H_x) = r + sH_x \quad (2),$$

where $r$ and $s$ are fitting parameters. Figures 2(c) and 2(d) depict $V_{2f}$ under $H_x$ sweep for $d = 0$ and $d = 10 \times 10^{16}$ ions/cm², respectively, for different $I_{ac}$. With increase in $I_{ac}$, the slope $s$ is also found to increase. From the fitting parameters, the damping-like effective field ($H_{DL}$) can be calculated according to the relation,[37]

$$H_{DL} = -2\left[\frac{\left(\frac{\partial V_{2f}}{\partial H_x}\right)}{\left(\frac{\partial^2 V_{1f}}{\partial H_x^2}\right)}\right] = -\frac{s}{c} \quad (3).$$

To determine $\theta_{SH}$, $H_{DL}$ was evaluated for various values of $I_{ac}$. Assuming a parallel resistor model and using the measured resistivities of the various layers, namely, Pt (= 27 µΩ cm), Pt (P) $5 \times 10^{16}$ (= 306 µΩ cm), Pt (P) $10 \times 10^{16}$ (= 882 µΩ cm) and GdFe (= 346 µΩ cm), the current density in the NM ($J_{NM}$) was estimated. Figure 3 (a) shows $H_{DL}$ as a function of $J_{NM}$ for different $d$ values. Performing a linear fit on this data, the ratio of $H_{DL}$ to $J_{NM}$ was calculated, and substituted in,[38]

$$\theta_{SH} = \frac{2e\mu_0 M_s t_{NM}}{\hbar}\left(\frac{H_{DL}}{J_{NM}}\right) \quad (4),$$

where $\mu_0$ is the vacuum permeability, $M_S$ is the saturation magnetization (= 47 mT), and $t_{NM}$ is the thickness of the NM layer. The calculated $\theta_{SH}$ as a function of $d$ is given in Fig. 3(b). $\theta_{SH}$ was found to increase from 0.06 for $d = 0$, to 0.43 for $d = 10 \times 10^{16}$ ions/cm², increasing by a factor of about 7.

In order to investigate the direct effect of this improved $\theta_{SH}$ on perpendicular magnetization switching, we proceeded to perform SOT driven switching measurements. Square write current $I_{write}$ pulses having pulse width $t_{pw} = 500$ μs were given as input. Figures 4(a) and 4 (b) depict normalized $V_{Hall}$ as a function of write current density in the NM layer $J_{write}$, in the presence of assisting fields of + 50 mT and − 50 mT, respectively, for different $d$. The sense of switching reverses with reversal of field polarity. + (−) $H_x$ results in a(n) (anti-)clockwise sense of switching. Complete and reversible switching of the perpendicular magnetization was observed irrespective of $d$ (S1 in the supplementary material). We define $J_{sw}$ as the current density at which $V_{Hall}$ changes polarity. We proceeded to evaluate current driven switching with different values of assisting $H_x$. $J_{sw}$ obtained as a function of $H_x$ for the various $d$ values are given in Fig. 4(c). For a fixed $H_x$, $J_{sw}$ is found to decrease with increasing $d$. For $d = 0$, $J_{sw}$ is found to be in the range of $4.0 \times 10^{11}$ A/m². However, for d = $5 \times 10^{16}$ and $10 \times 10^{16}$ ions/cm², $J_{sw}$ decreases to the range of $3.0 \times 10^{10}$ A/m² and $1.5 \times 10^{10}$ A/m², respectively. This trend may be understood to arise from a combination of two factors: the decrease in perpendicular anisotropy for nonzero $d$ (Fig. 1(d)) and the increase in $\theta_{SH}$ with increasing $d$ (Fig. 3(b)). The write energy densities for different $d$ were estimated, and found to reduce by about a factor of 25 when $d$ was increased from zero to $10 \times 10^{16}$ ions/cm² (S2 in the supplementary material).

Further, to reduce the thermal assistance to switching, we performed switching using pulses of shorter $t_{pw}$ ranging from 250 ns to 500 μs, on the samples with nonzero $d$. The measurements were performed under an assist field of + 100 mT. $J_{sw}$ is given as a function of $t_{pw}$, for both samples with $d = 5 \times 10^{16}$ and $10 \times 10^{16}$ ions/cm², in Fig. 5 (a). $J_{sw}$ is found to increase with decrease in $t_{pw}$ for both values of $d$, with lower $J_{sw}$ for higher $d$ value. When $t_{pw} < 1$ μs (intrinsic regime), efficient angular momentum transfer to the magnetization determines the switching speed and $J_{sw}$ varies drastically with $t_{pw}$,[39] as seen in Fig. 5 (a). Meanwhile, for $t_{pw} > 1$ μs (thermally activated regime), thermal fluctuations offer additional assistance to switching and $J_{sw}$ varies weakly with $t_{pw}$.[39] The data of Fig. 5 (a) in the thermally activated regime is fitted to the expression,[40]

$$J_{sw}(t_{pw}) = J_{sw\ 0}\left[1 - \frac{1}{\Delta}\ln\left(\frac{t_{pw}}{t_0}\right)\right] \quad (5),$$

where $J_{sw\ 0}$ is the switching current density at zero thermal fluctuation and $t_0$ is the thermal fluctuation time (typically 1 ns).

$J_{sw\ 0}$ and $\Delta$ obtained for different $d$ are shown in Fig. 5 (b). $J_{sw\ 0}$ is found to decrease by a factor of 4, from $8.4 \times 10^{10}$ A/m² for $d = 5 \times 10^{16}$ ions/cm² to $2.0 \times 10^{10}$ A/m² for $d = 10 \times 10^{16}$ ions/cm². However, $\Delta$ is found to change negligibly with increasing $d$, going from 62.6 for $d = 5 \times 10^{16}$ ions/cm² to 61.4 for $d = 10 \times 10^{16}$ ions/cm². This indicates that the contribution of the reduced $H_p$ to decreasing $J_{sw}$ is negligible in comparison to the observed increase in $\theta_{SH}$. Further, $\Delta$ is found to be nearly invariant under P implantation doses of upto $10 \times 10^{16}$ ions/cm². Thus, using Pt (P) $d$ as the SOT source can significantly reduce the critical switching current density required to switch a perpendicular magnetization (by a factor of nearly 27 in this work), with minimal change in thermal stability, while maintaining perpendicular magnetic anisotropy in the samples. Further, we proceeded to calculate $\theta_{SH}$ using the values of $J_{sw\ 0}$ and $H_c$ estimated previously, according to the expression,[41]

$$\theta_{SH} = \left(\frac{4}{\pi}\right)\left(\frac{e\mu_0 M_s t_{FiM}}{\hbar}\right)\left(\frac{H_c}{J_{sw\ 0}}\right) \qquad (6),$$

where $t_{FiM}$ denotes the thickness of the FiM layer. $\theta_{SH}$ is found to increase from 0.12 for $d = 5 \times 10^{16}$ ions/cm² to 0.43 for $d = 10 \times 10^{16}$ ions/cm², as shown in Fig. 5 (c). These values agree with the trend observed from harmonic Hall results in Fig. 3 (b).

In conclusion, we showed improved charge-to-spin conversion efficiency, quantified in terms of $\theta_{SH}$, when using phosphorus implanted Pt, Pt (P) as the NM in subs./NM/Pt/GdFe/SiN stacks, with GdFe possessing PMA. By the harmonic Hall technique, $\theta_{SH}$ was found to increase from 0.06 in Pt to 0.19 in Pt with P implantation dose $d = 5 \times 10^{16}$ ions/cm². For higher $d$ of $10 \times 10^{16}$ ions/cm², $\theta_{SH}$ was further enhanced to 0.43. Thus, overall $\theta_{SH}$ was found to increase seven times when using P implanted Pt as the NM. The effect of this improved $\theta_{SH}$ in perpendicular magnetization switching was investigated and found to decrease $J_{sw}$ to a value of 27 times smaller than that obtained for Pt. The choice of Pt (P) as the NM was also found to affect the perpendicular anisotropy of GdFe, reducing $H_p$ for $d = 10 \times 10^{16}$ ions/cm² to nearly one-third of the corresponding value for $d = 0$. Further, the order of decrease in $J_{sw}$ was confirmed to be maintained even after accounting for the thermal effects, via pulse width dependent switching measurements. Thermal stability was found to change negligibly for different $d$, remaining sufficiently above the 10 year data retention benchmark of 45.[42] Thus, NM/GdFe stacks consisting of Pt (P) with enhanced $\theta_{SH}$, having reduced perpendicular anisotropy, sufficient thermal stability and significant energy efficiency could be promising candidates for use in next generation SOT-based MRAMs and compute-in-memory architectures.

**Supplementary Material**

See the supplementary material for additional details regarding complete perpendicular magnetization switching and switching write energy densities for different phosphorus implantation doses.

**Acknowledgements**

This work was partially supported by JSPS Grant-in-Aid (KAKENHI No. 22K04198), Iketani Science and Technology Foundation, Heiwa Nakajima Foundation and Meisenkai.

**Author Declarations**

**Conflict of Interest**

The authors have no conflicts of interest to declare.

**Author Contributions**

Kazuki Shintaku and Arun Jacob Mathew contributed equally to this work.

**Kazuki Shintaku:** Data curation (equal); Formal analysis (equal); Validation (equal); Writing – original draft (equal), **Arun Jacob Mathew**: Conceptualization (equal); Investigation (equal); Data curation (equal); Formal analysis (equal); Validation (equal); Writing – original draft (equal), **Akihisa Iwamoto:** Data curation (equal); Formal analysis (equal); Validation (equal), **Mojtaba Mohammadi:** Supervision (equal); Writing – review & editing (equal), **Hiroyuki Awano:** Resources (equal); Supervision (equal); Writing – review & editing (equal), **Hironori Asada:** Funding acquisition (equal); Resources (equal); Supervision (equal); Writing – review & editing (equal), **Yasuhiro Fukuma:** Funding acquisition (equal); Project administration (equal); Resources (equal); Conceptualization (equal); Supervision (equal); Visualization (equal); Writing – review & editing (equal).

**Data availability**

The data that supports the findings of this study are available from the corresponding author upon reasonable request.

**References**:

[1] L. Berger, "Emission of spin waves by a magnetic multilayer traversed by a current," Phys Rev B **54**(13), 9353–9358 (1996).

[2] J.C. Slonczewski, "Current-driven excitation of magnetic multilayers," J Magn Magn Mater **159**(1–2), L1–L7 (1996).

[3] S. Ikegawa, F.B. Mancoff, and S. Aggarwal, "Commercialization of MRAM – Historical and Future Perspective," in *2021 IEEE International Interconnect Technology Conference (IITC)*, (IEEE, 2021), pp. 1–3.

[4] D.C. Ralph, and M.D. Stiles, "Spin transfer torques," J Magn Magn Mater **320**(7), 1190–1216 (2008).

[5] J.H. Lim, N. Raghavan, A. Padovani, J.H. Kwon, K. Yamane, H. Yang, V.B. Naik, L. Larcher, K.H. Lee, and K.L. Pey, "Investigating the Statistical-Physical Nature of MgO Dielectric Breakdown in STT-MRAM at Different Operating Conditions," in *2018 IEEE International Electron Devices Meeting (IEDM)*, (IEEE, 2018), pp. 25.3.1-25.3.4.

[6] I.M. Miron, K. Garello, G. Gaudin, P.-J. Zermatten, M. V. Costache, S. Auffret, S. Bandiera, B. Rodmacq, A. Schuhl, and P. Gambardella, "Perpendicular switching of a single ferromagnetic layer induced by in-plane current injection," Nature **476**(7359), 189–193 (2011).

[7] L. Liu, O.J. Lee, T.J. Gudmundsen, D.C. Ralph, and R.A. Buhrman, "Current-Induced Switching of Perpendicularly Magnetized Magnetic Layers Using Spin Torque from the Spin Hall Effect," Phys Rev Lett **109**(9), 096602 (2012).

[8] P. Gambardella, and I.M. Miron, "Current-induced spin–orbit torques," Philosophical Transactions of the Royal Society A: Mathematical, Physical and Engineering Sciences **369**(1948), 3175–3197 (2011).

[9] A. Manchon, J. Železný, I.M. Miron, T. Jungwirth, J. Sinova, A. Thiaville, K. Garello, and P. Gambardella, "Current-induced spin-orbit torques in ferromagnetic and antiferromagnetic systems," Rev Mod Phys **91**(3), 035004 (2019).

[10] M. Cubukcu, O. Boulle, N. Mikuszeit, C. Hamelin, T. Bracher, N. Lamard, M.-C. Cyrille, L. Buda-Prejbeanu, K. Garello, I.M. Miron, O. Klein, G. de Loubens, V. V. Naletov, J. Langer, B. Ocker, P. Gambardella, and G. Gaudin, "Ultra-Fast Perpendicular Spin–Orbit Torque MRAM," IEEE Trans Magn **54**(4), 1–4 (2018).

[11] Z. Guo, J. Yin, Y. Bai, D. Zhu, K. Shi, G. Wang, K. Cao, and W. Zhao, "Spintronics for Energy- Efficient Computing: An Overview and Outlook," Proceedings of the IEEE **109**(8), 1398–1417 (2021).

[12] J.R. Mohan, A.J. Mathew, K. Nishimura, R. Feng, R. Medwal, S. Gupta, R.S. Rawat, and Y. Fukuma, "Classification tasks using input driven nonlinear magnetization dynamics in spin Hall oscillator," Sci Rep **13**(1), 7909 (2023).

[13] A. Jacob Mathew, Y. Gao, J. Wang, M. Mohammadi, H. Awano, M. Takezawa, H. Asada, and Y. Fukuma, "Realization of logic operations via spin–orbit torque driven perpendicular magnetization switching in a heavy metal/ferrimagnet bilayer," J Appl Phys **135**(21), 213902 (2024).

[14] V.M. Edelstein, "Spin polarization of conduction electrons induced by electric current in two-dimensional asymmetric electron systems," Solid State Commun **73**(3), 233–235 (1990).

[15] J.E. Hirsch, "Spin Hall Effect," Phys Rev Lett **83**(9), 1834–1837 (1999).

[16] T. Suzuki, S. Fukami, N. Ishiwata, M. Yamanouchi, S. Ikeda, N. Kasai, and H. Ohno, "Current-induced effective field in perpendicularly magnetized Ta/CoFeB/MgO wire," Appl Phys Lett **98**(14), 142505 (2011).

[17] G. Yu, P. Upadhyaya, K.L. Wong, W. Jiang, J.G. Alzate, J. Tang, P.K. Amiri, and K.L. Wang, "Magnetization switching through spin-Hall-effect-induced chiral domain wall propagation," Phys Rev B **89**(10), 104421 (2014).

[18] S.K. Kim, G.S.D. Beach, K.-J. Lee, T. Ono, T. Rasing, and H. Yang, "Ferrimagnetic spintronics," Nat Mater **21**(1), 24–34 (2022).

[19] Z. Zhao, M. Jamali, A. K. Smith, and J.-P. Wang, "Spin Hall switching of the magnetization in Ta/TbFeCo structures with bulk perpendicular anisotropy," Appl Phys Lett **106**(13), 132404 (2015).

[20] N. Roschewsky, T. Matsumura, S. Cheema, F. Hellman, T. Kato, S. Iwata, and S. Salahuddin, "Spin-orbit torques in ferrimagnetic GdFeCo alloys," Appl Phys Lett **109**(11), 112403 (2016).

[21] J. Finley, and L. Liu, "Spin-Orbit-Torque Efficiency in Compensated Ferrimagnetic Cobalt-Terbium Alloys," Phys Rev Appl **6**(5), 054001 (2016).

[22] R. Mishra, J. Yu, X. Qiu, M. Motapothula, T. Venkatesan, and H. Yang, "Anomalous Current-Induced Spin Torques in Ferrimagnets near Compensation," Phys Rev Lett **118**(16), 167201 (2017).

[23] K.-J. Kim, S.K. Kim, Y. Hirata, S.-H. Oh, T. Tono, D.-H. Kim, T. Okuno, W.S. Ham, S. Kim, G. Go, Y. Tserkovnyak, A. Tsukamoto, T. Moriyama, K.-J. Lee, and T. Ono, "Fast domain wall motion in the vicinity of the angular momentum compensation temperature of ferrimagnets," Nat Mater **16**(12), 1187–1192 (2017).

[24] S. Ranjbar, S. Kambe, S. Sumi, P. V. Thach, Y. Nakatani, K. Tanabe, and H. Awano, "Elucidation of the mechanism for maintaining ultrafast domain wall mobility over a wide temperature range," Mater Adv **3**(18), 7028–7036 (2022).

[25] M. Mohammadi, S. Sumi, K. Tanabe, and H. Awano, "Boosting domain wall velocity and stability in ferrimagnetic GdFe nanowires by applying laser-annealing process," Appl Phys Lett **123**(20), 202403 (2023).

[26] M. Mohammadi, S. Ranjbar, P. Van Thach, S. Sumi, K. Tanabe, and H. Awano, "Exploring fast domain wall motion and DMI realization in compensated ferrimagnetic nanowires," J Phys D Appl Phys **58**(5), 055002 (2025).

[27] U. Shashank, R. Medwal, T. Shibata, R. Nongjai, J.V. Vas, M. Duchamp, K. Asokan, R.S. Rawat, H. Asada, S. Gupta, and Y. Fukuma, "Enhanced Spin Hall Effect in S-Implanted Pt," Adv Quantum Technol **4**(1), 2000112 (2021).

[28] U. Shashank, Y. Nakamura, Y. Kusaba, T. Tomoda, R. Nongjai, A. Kandasami, R. Medwal, R.S. Rawat, H. Asada, S. Gupta, and Y. Fukuma, "Disentanglement of intrinsic

and extrinsic side-jump scattering induced spin Hall effect in N-implanted Pt," Phys Rev B **107**(6), 064402 (2023).

[29] U. Shashank, R. Medwal, Y. Nakamura, J.R. Mohan, R. Nongjai, A. Kandasami, R.S. Rawat, H. Asada, S. Gupta, and Y. Fukuma, "Highly dose dependent damping-like spin–orbit torque efficiency in O-implanted Pt," Appl Phys Lett **118**(25), 252406 (2021).

[30] U. Shashank, Y. Kusaba, J. Nakamura, A.J. Mathew, K. Imai, S. Senba, H. Asada, and Y. Fukuma, "Charge–spin interconversion in nitrogen sputtered Pt via extrinsic spin Hall effect," Journal of Physics: Condensed Matter **36**(32), 325802 (2024).

[31] E. Lesne, Y. Fu, S. Oyarzun, J.C. Rojas-Sánchez, D.C. Vaz, H. Naganuma, G. Sicoli, J.-P. Attané, M. Jamet, E. Jacquet, J.-M. George, A. Barthélémy, H. Jaffrès, A. Fert, M. Bibes, and L. Vila, "Highly efficient and tunable spin-to-charge conversion through Rashba coupling at oxide interfaces," Nat Mater **15**(12), 1261–1266 (2016).

[32] Z. Wang, D. Ki, H. Chen, H. Berger, A.H. MacDonald, and A.F. Morpurgo, "Strong interface-induced spin–orbit interaction in graphene on WS2," Nat Commun **6**(1), 8339 (2015).

[33] K. Kondou, R. Yoshimi, A. Tsukazaki, Y. Fukuma, J. Matsuno, K.S. Takahashi, M. Kawasaki, Y. Tokura, and Y. Otani, "Fermi-level-dependent charge-to-spin current conversion by Dirac surface states of topological insulators," Nat Phys **12**(11), 1027–1031 (2016).

[34] H. Wu, Y. Xu, P. Deng, Q. Pan, S. A. Razavi, K. Wong, L. Huang, B. Dai, Q. Shao, G. Yu, X. Han, J.C. Rojas-Sánchez, S. Mangin, and K. L. Wang, "Spin-Orbit Torque

Switching of a Nearly Compensated Ferrimagnet by Topological Surface States," Advanced Materials **31**(35), 1901681 (2019).

[35] N. Nagaosa, J. Sinova, S. Onoda, A.H. MacDonald, and N.P. Ong, "Anomalous Hall effect," Rev Mod Phys **82**(2), 1539–1592 (2010).

[36] U.H. Pi, K. Won Kim, J.Y. Bae, S.C. Lee, Y.J. Cho, K.S. Kim, and S. Seo, "Tilting of the spin orientation induced by Rashba effect in ferromagnetic metal layer," Appl Phys Lett **97**(16), 162507 (2010).

[37] J. Kim, J. Sinha, M. Hayashi, M. Yamanouchi, S. Fukami, T. Suzuki, S. Mitani, and H. Ohno, "Layer thickness dependence of the current-induced effective field vector in Ta|CoFeB|MgO," Nat Mater **12**(3), 240–245 (2013).

[38] M.H. Nguyen, and C.F. Pai, "Spin-orbit torque characterization in a nutshell," APL Mater **9**(3), 030902 (2021).

[39] K. Garello, C.O. Avci, I.M. Miron, M. Baumgartner, A. Ghosh, S. Auffret, O. Boulle, G. Gaudin, and P. Gambardella, "Ultrafast magnetization switching by spin-orbit torques," Appl Phys Lett **105**(21), 212402 (2014).

[40] E.B. Myers, F.J. Albert, J.C. Sankey, E. Bonet, R.A. Buhrman, and D.C. Ralph, "Thermally Activated Magnetic Reversal Induced by a Spin-Polarized Current," Phys Rev Lett **89**(19), 196801 (2002).

[41] A. Thiaville, S. Rohart, É. Jué, V. Cros, and A. Fert, "Dynamics of Dzyaloshinskii domain walls in ultrathin magnetic films," EPL (Europhysics Letters) **100**(5), 57002 (2012).

[42] B. Dieny, and M. Chshiev, "Perpendicular magnetic anisotropy at transition metal/oxide interfaces and applications," Rev Mod Phys **89**(2), 025008 (2017).

**Figures:**

**Figure 1**:
(a) Schematic of the setup for measuring the Hall voltage $V_{Hall}$. (b) Normalized $V_{Hall}$ under out-of-plane field ($H_z$) sweep. (c) Normalized $V_{Hall}$ under in-plane field ($H_x$) sweep. The solid lines indicate linear fits to the data. (d) Coercivity under $H_z$ sweep, $H_c$ and perpendicular anisotropy field $H_p$ as functions of phosphorus implantation dose $d$.

**Figure 2:**
(a) First harmonic Hall voltage $V_{1f}$ as a function of $H_x$ for different input current amplitudes $I_{ac}$ ($d = 0$). (b) $V_{1f}$ as a function of $H_x$ for different $I_{ac}$ ($d = 10 \times 10^{16}$ ions/cm$^2$). (c) Second harmonic Hall voltage $V_{2f}$ as a function of $H_x$ for different $I_{ac}$ ($d = 0$). (d) $V_{2f}$ as a function of $H_x$ for different $I_{ac}$ ($d = 10 \times 10^{16}$ ions/cm$^2$).

**Figure 3:**
(a) The effective damping-like field $H_{DL}$ as a function of current density through the nonmagnet layer $J_{NM}$. The spin Hall angle $\theta_{SH}$ is obtained from the slope of each data set. (b) $\theta_{SH}$ as functions of dose $d$. With increase in $d$, monotonic increase in $\theta_{SH}$ is observed.

**Figure 4:**
Normalized $V_{Hall}$ as a function of write current density $J_{write}$ in the presence of an assisting field $H_x$ of (a) + 50 mT and (b) − 50 mT for different $d$. (c) Critical switching current density $J_{sw}$ as a function of $H_x$ for different $d$.

**Figure 5:**
(a) $J_{sw}$ as a function of pulse width $t_{pw}$ for different $d$. (b) Critical switching current density at zero thermal fluctuations $J_{sw\ 0}$ and thermal stability factor $\Delta$ as functions of different $d$. (c) $\theta_{SH}$, estimated from pulse width dependent switching measurements, for different $d$.

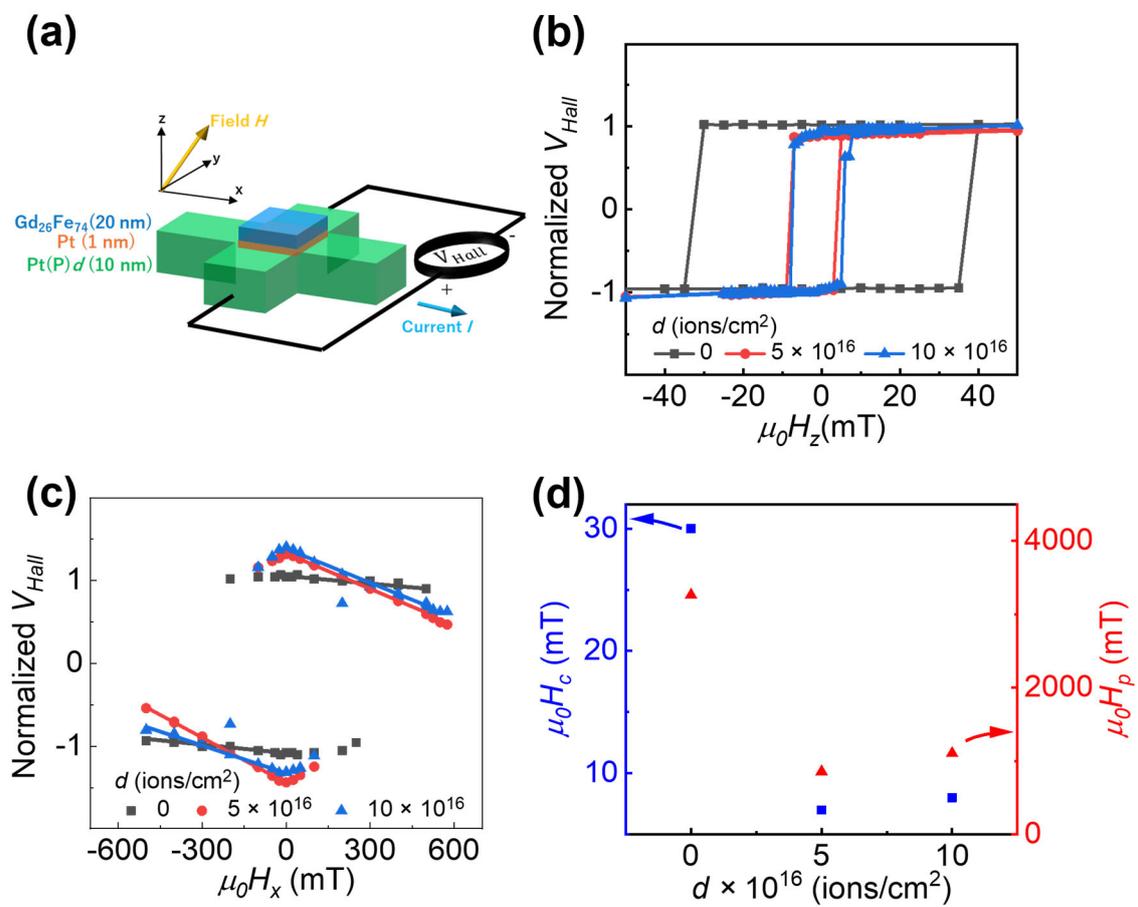

Figure 1, K. Shintaku *et al.*

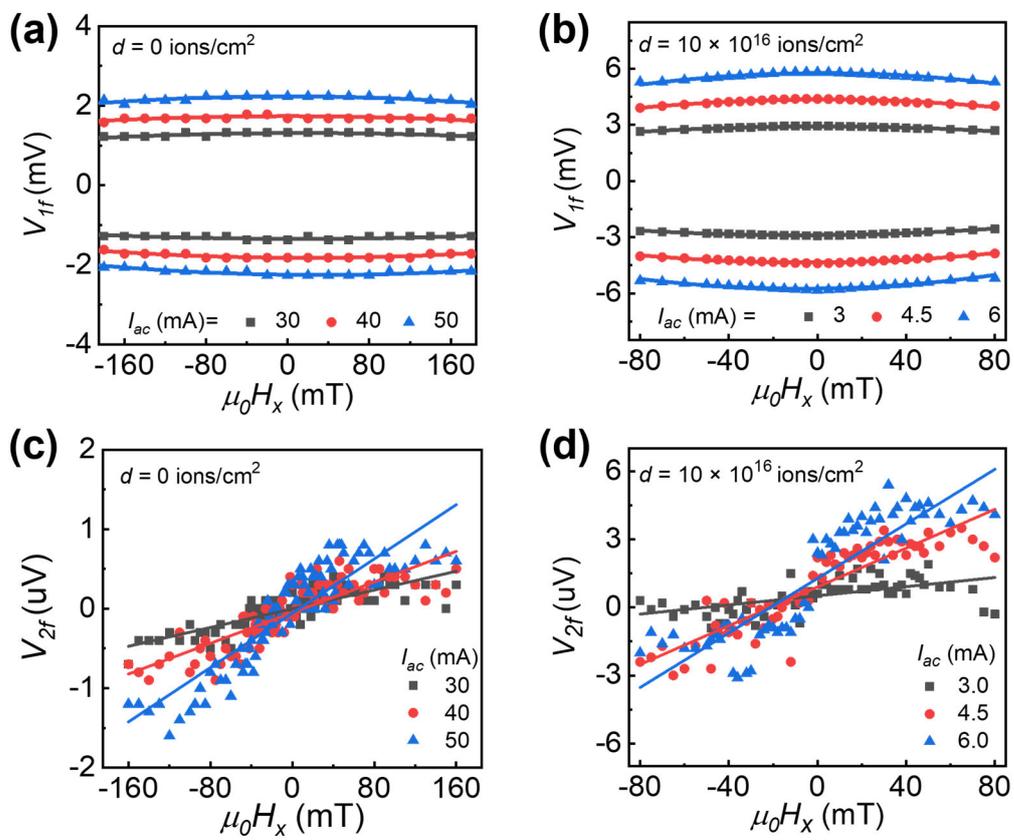

Figure 2, K. Shintaku *et al.*

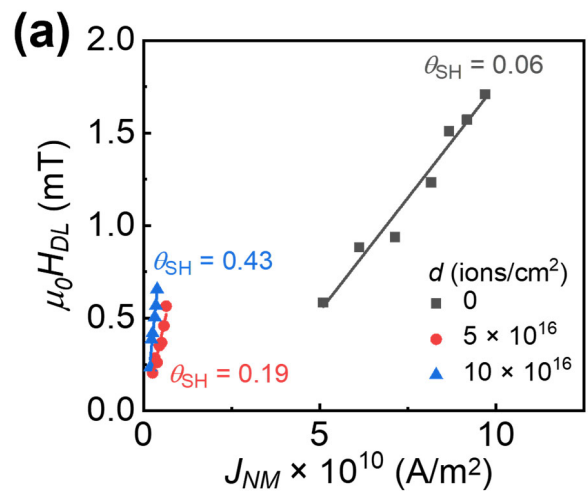

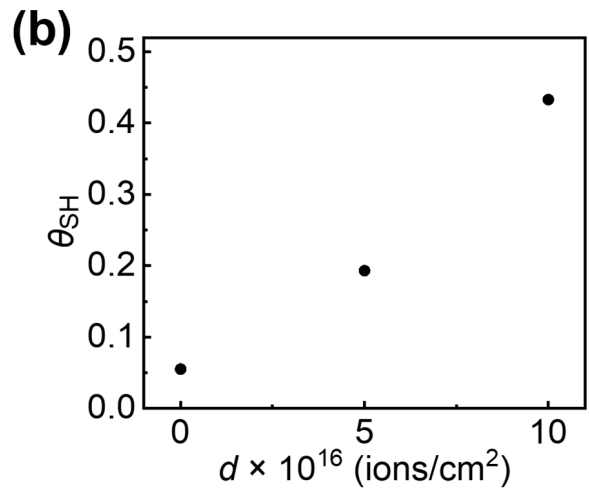

Figure 3, K. Shintaku *et al*

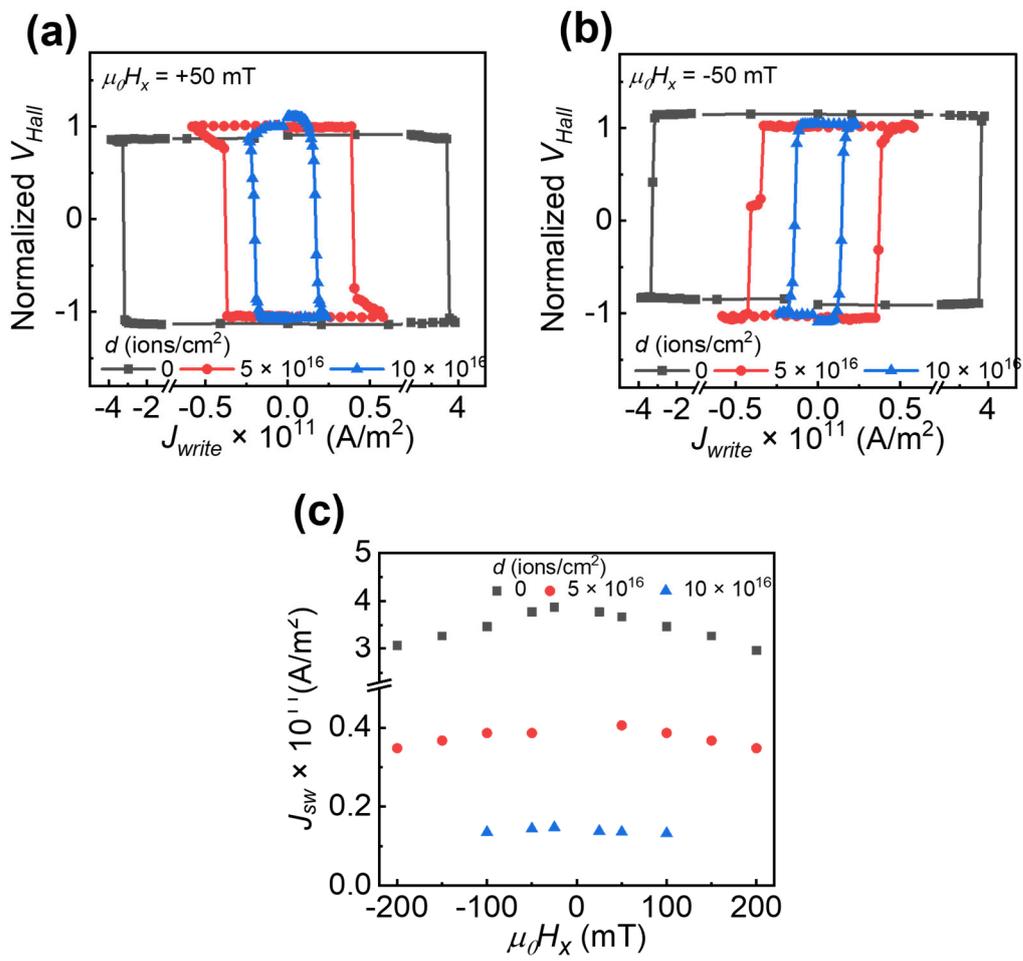

Figure 4, K. Shintaku *et al.*

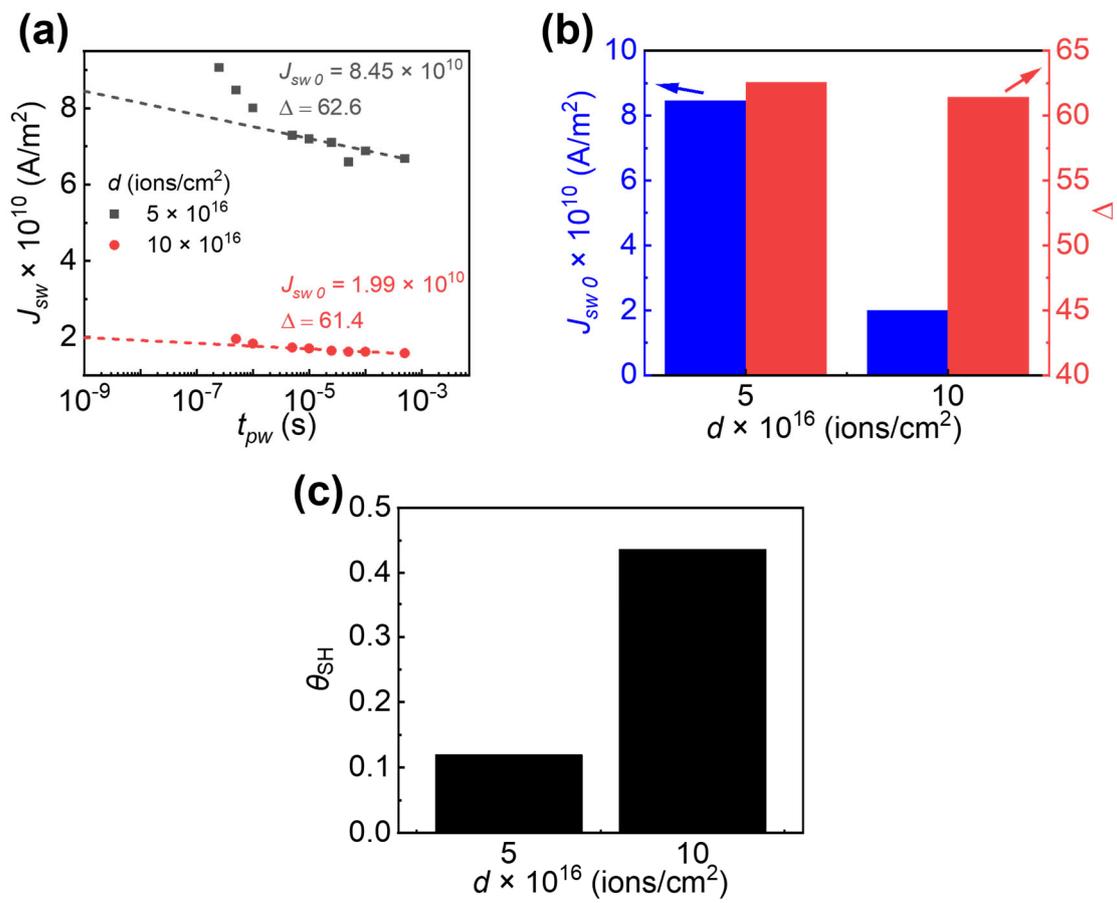

Figure 5, K. Shintaku *et al.*